\documentstyle[aps,pre,epsfig,graphicx]{revtex}
\newcommand{\be}{\begin{equation}}
\newcommand{\bea}{\begin{eqnarray} \nonumber}
\newcommand{\ee}{\end{equation}}
\newcommand{\eea}{\end{eqnarray}}

 \def\(({\left(}
 \def\)){\right)}
\def\[[{\left[}
\def\]]{\right]}
\def\bi{\bibitem}
\def \form#1 {eq. (\ref{#1}) }
\def \parziale#1#2  {{\partial {#1} \over \partial {#2}}}

\def \ba#1 {\overline{#1}}

\begin{document}

\title{Euclidean random matrices, the glass transition and the Boson peak}

\author{Giorgio Parisi}
\address{Dipartimento di Fisica, Sezione INFN, SMC and UdRm1 of INFM,\\
Universit\`a di Roma ``La Sapienza'',
Piazzale Aldo Moro 2,
I-00185 Rome (Italy)\\
giorgio.parisi@roma1.infn.it}

\date{\today}

\maketitle
\begin{abstract}
In this paper I will describe some results that have been recently obtained in the study of 
random Euclidean matrices, i.e. matrices that are functions of random points in Euclidean 
space. In the case of {\sl translation invariant} matrices one generically finds a phase transition 
between a {\sl phonon} phase and a {\sl saddle} phase.  If we apply these considerations 
to the study of the Hessian of the Hamiltonian of the particles of a fluid, we find that this 
phonon-saddle transition corresponds to the dynamical phase transition in glasses, that 
has been studied in the framework of the mode coupling approximation. The Boson peak 
observed in glasses at low temperature is a remanent of this transition.
\end{abstract}

\section{Introduction}

In the last years many people have worked on the problem of computing the properties of Euclidean 
random matrices.  The problem can be formulated as follows.  We consider a set of $N$ 
points ($x_{i}$) that are randomly distributed with some given distribution (two extreme 
examples are: a) the $x$'s are random independent points, b) the $x$'s are one of the many 
minima of a given Hamiltonian). 

In the simplest formulation, given a function $h(x)$, we 
consider the $N \times N $ matrix:
\be
M_{i,k}=h(x_{i}-x_{k}) \ .
\ee
The problems consists in computing the properties of the eigenvalues and of the 
eigenvectors of $M$.  Of course, for finite $N$ they will depend on the instance of the 
problem (i.e. the actually choice of the $x$'s), however system to system fluctuations 
for the intensive quantities (e.g. the spectral density $\rho(z)$) disappear when we 
consider the limit $ N \to \infty$ at fixed particle density.

The problem is not new; it has been carefully studied in the case where the positions of 
the particles are restricted to be on the lattice.  The case where the 
particles can stay in arbitrary positions, that is relevant in the study of fluids, has 
been less studied, although in the past many papers have been written on the argument 
\cite{MOLTE}. These off-lattice models present some technical (and also physical) differences with 
on-lattice models.

A certain number of variations are interesting. For example we could consider the case 
where we add a term on the diagonal:
\be
M_{i,k}=\delta_{i,k} \sum_{j}h(x_{i}-x_{j}) -h(x_{i}-x_{k}) \ .
\ee
In this case  the associated quadratic form is given by
\be
\sum_{i,k}\phi_{i}\phi_{k}M_{i,k}=\frac12 \sum_{i,k}h(x_{i}-x_{k}) ( \phi_{i}-\phi_{k})^{2}
\ee
and therefore the matrix $M$ is non-negative if the function $h$ is non-negative.  The 
matrix $M$ has always a zero eigenvalue as consequence of the invariance of the quadratic 
form under the symmetry $\phi_{i} \to \phi_{i} +\lambda$; the presence of this symmetry has 
deep consequences on the properties of the spectrum of $M$ and, as we shall see later, {\sl phonons} 
are present.

In the same spirit we can consider a two-body potential $V(x)$ and introduce the 
Hamiltonian 
\be
H[x]=\frac12 \sum_{i,k}V(x_{i}-x_{k}) \ .
\ee
We can consider the $3N \times 3N $ Hessian matrix
\be
M_{i,k}=
\delta_{i,k} \sum_{j}V''(x_{i}-x_{j}) -V''(x_{i}-x_{k}) \ .\label{HESSIAN}
\ee
where for simplicity we have not indicated space indices.
Also here we are interested in the computation of the spectrum of $M$.  The translational 
invariance of the Hamiltonian implies that the associated quadratic form is invariant 
under the symmetry $\phi_{i} \to \phi_{i} +\lambda$ and a phonon-like behaviour may be 
expected.

This tensorial case, especially when the distribution of the $x$'s is related to the potential 
$V$, is the most interesting from the physical point of view (especially for its 
relations to the theory of fluids), however 
it is technically more involved, so that it may be convenient to study in details the 
previous scalar cases in order to improve our command of the tools that are needed in the 
computations.

Our aim is to get both a qualitative understanding of the main properties of the spectrum 
and of the eigenvalues and a quantitate, as accurate as possible, analytic evaluation of 
these properties. Quantitative accurate results are also needed because the computation of 
the spectral density is a crucial step in the microscopic computation of the thermodynamic 
dynamic properties of glass forming systems. In some sense this approach can be considered 
as an alternative route to obtain mode-coupling like result, the main difference being 
that in the mode coupling  one uses a coarse grained approach and  the 
hydrodynamical equations, while here we take a fully microscopic point of view.

Before describing the progresses that have been done in the field I will recall some of 
the physical interesting cases for which this approach may be useful. This will done in 
the next two sections.

\section{Glasses at low temperature}

Let us consider a  potential $V$ such that the system forms a glass at low temperature.
In this situation the Hamiltonian $H$ has an exponential large number of minima 
i.e. proportional to $\exp(a N)$ where $a$ is a number of order 1. We can consider two 
different cases: (a)  the variables $x$ are at the absolute minimum of $H$,  (b) the 
variables $x$ belong to a generic minimum of energy density $E$.

In the real world, if we cool a glass at very low temperatures, it will not evolve toward 
the absolute minimum, characterized by an energy density $E_{0}$, but it will evolve 
toward local minima that have an higher energy density $E$, depending on the details of 
the history of the sample; these minima are usually called inherent structures.  In both 
cases, at low temperature the dynamics and the thermodynamics will be dominated by small 
harmonic vibrations around the minimum, i.e. the ISNM (inherent structure normal modes).

Qualitatively the spectrum of these vibrations is well known.  No negative eigenvalues are 
present; at very low energy we have the phonons, at higher energy we have the famous Boson 
peak (the Boson peak is defined as a bump at some small but non-zero value of the 
eigenvalue $z$ of the spectral density $\rho(z)$ divided $z^{2}$), \footnote{The ratio 
$\rho(z)/z^{1/2}$ usually go to a constant at $z=0$, if we consider the the spectrum of the 
vibrations of a solid; at low value of the eigenvalue $z$ the vibrations are phonons.} 
whose origine will be discussed later in details, at still higher energy we have the main 
component of the spectrum and at very high energy there is the tail of localized states.  
The presence of some localized states in the low energy region is still matter of debate.  
The fact that the sound velocity is approximatively constant in the region of the Boson 
peak was considered amazing.

There are many other quantities that we would like to compute, in particular the structure 
function $S(z,k)$ that can be directly experimentally measured. Moreover the width 
of the acoustic peak $\Gamma(z)$ has an interesting behaviour at low $z$ that is 
related to the behaviour of sound attenuation as function of the frequency.

In the long run we are also interested to understand quantum effects.  There are {\sl 
trivial} quantum effects that are related to the form of the spectral density; there are 
also more interesting quantum effects, e.g. two levels systems, that cannot be explained 
in the small fluctuation limit. However it is quite plausible the a good control of the 
quadratic part of the potential may be useful in getting quantitative predictions on these 
non linear effects.

\section{The dynamical glass transition}
\subsection{The mean field scenario}
It is usually believed that in the real world fragile glasses have only one transition 
with divergent viscosity (at temperature $T_{K}$).  This transition cannot be observed 
directly because the time required for thermalization is too long.  This transition is 
believed to be related to Kauzmann entropy crisis and it should happens just at the point 
where the configurational entropy becomes zero.  The viscosity is supposed to diverge a 
$\exp(A/(T-T_{K}))$ and the specific heat should be discontinuous.

However in the idealized world of mean field theories (e.g. the mode coupling theory or 
the equivalent replica approach), where activated processes are strictly forbidden, there 
is a second purely dynamics transition $T_{c}$ at higher temperatures \cite{tirumma}.  Here the viscosity 
is divergent as a power law of $T-T_{c}$.  This idealization is not so bad in the real 
world: activated processes are strongly depressed and the viscosity may increase of many order 
of magnitudo (e.g. 6) before reaching the region where activated processes become dominant.

As it happens in many cases, slow relaxation is related to the existence of zero energy 
modes and this statement is true also in the mode coupling theory. This statement can be 
easily verified in spin models where the mode coupling theory is exact and simple 
computations are possible.
\begin{center} \begin{figure}[bt] {\epsfysize=8cm       \epsffile{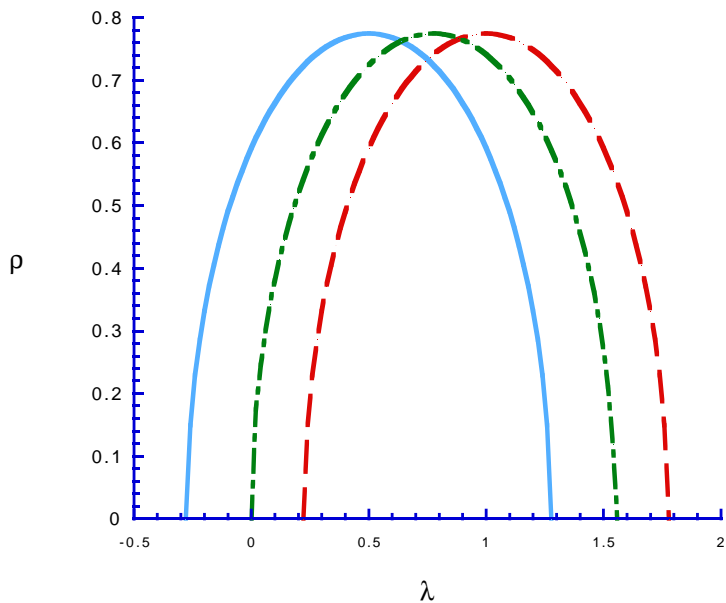}
    \caption{The qualitative behaviour of the spectrum in mean field approximation above 
    $T_{c}$, (full line), at $T_{c}$ (dot-dashed line) and below $T_{c}$ (dashed line) as 
    function of the eigenvalue $\lambda$).
    }\label{X}
    }\end{figure}\end{center}

The general idea is quite simple.  Let us consider the free energy as functional of the 
density $\rho(x)$ (i.e. $F[\rho]$) (in a magnetic system the free energy as function of 
all the local magnetizations).  We expect that at $T>T_{c}$ the only relevant solution for 
the thermodynamics is the trivial solution $\rho(x)=const$.  At low temperatures there are 
an exponential large number of non-equivalent solutions where the the density has a non 
trivial dependence on $x$.  Skipping many details the situation is the 
following:

\begin{itemize}
    \item A temperature $T>T_{c}$ there are no non-trivial relevant solutions of the equation 
    \be
    {\delta F \over \delta \rho(x)}=0, \label {D1F}
    \ee
    however the dynamics is dominated by quasi solutions of the previous equations, i. e. 
    by densities $\rho(x)$ such that the left hand side of the previous equation is not 
    zero, but vanishes when $T$ approach $T_{c}$. These quasi solutions are relevant for 
    the dynamics \cite{FV}. One can compute the spectrum of the Hessian
    \be
    M(x,y)={\delta^{2} F \over \delta \rho(x) \delta \rho(y)}\label {D2F} \ .
    \ee
    One finds that  $M$ has negative eigenvalues and its spectrum extends to
    the negative eigenvalue region and has qualitatively the shape shown in fig. \ref{X}.
    These quasi stationary points of $F$ look like saddles.
    \item
    At the transition point $T=T_{c}$ the quasi stationary points becomes real solutions 
    of the equations (\ref{D1F}). They are essentially minima: the spectrum of the Hessian 
    is non-negative and it arrives up to zero. As it can be 
    checked directly, the existence of these nearly zero energy modes is responsible of 
    the slowing down of the dynamics. The different minima are connected by flat regions 
    so that the system may travel from one minimum to an other \cite{KL}.
    \item
    At low temperature the mimima become more deep, the spectrum develops a gap as shown 
    in fig. \ref{X} and the minima are no more connected by flat regions. If activated 
    processes were suppressed, the system would remains forever in one of these minima. In 
    the real world the system may jump  (by decreasing his energy) until it reaches the 
    region where the minima are so deep that the energy barriers among them diverges.
\end{itemize}

This picture is not so intuitive because it involves the presence of saddles with many 
directions in which the curvature is negative, and it is practically impossible to 
visualize it by making a drawing in a two or a three dimensional space.

This qualitative description can be easily verified in models where the mean field 
approximation is exact.  In glass forming liquid, the picture is essentially sound 
(provided that we correct it by considering the existence of phonons).  However, if we try 
to test it a more precise way, we face the difficulty that the free energy functional 
$F[\rho]$ is a mythological object whose exact form is not exactly known and consequently the 
eigenvalues of its Hessian cannot be computed.  We will see in the next two subsections 
that the so called instantaneous normal modes (INM) are a first approximation to the 
eigenvalues of the Hessian of $F$ and that a more interesting realization is provided by 
the saddle normal modes (SNM).

\subsection{Instantaneous normal modes}
Instantaneous normal modes in liquid and glasses have been studied for a long time.  They 
are defined as follows: we take a configuration at equilibrium at temperature $T$ (i.e. 
with a probability distribution proportional to $\exp(-\beta H))$ and for each 
configuration we compute the eigenvalues of the Hessian of the energy, given by 
eq. (\ref{HESSIAN}).

The behaviour of the fraction of INM with negative eigenvalues as the function of the 
temperature is both interesting and deceiving: interesting because this fraction seems to 
go to zero near $T_{c}$, if we look only to the data above $T_{c}$, deceiving because it 
remains distinctly different from zero also below $T_{c}$.

This is exactly what should happens: the spectrum of INM can be computed in mean field 
theory in simple models \cite{Biroli} and one finds systematically that the spectrum of 
INM has a part at negative energies both at $T_{c}$ and below $T_{c}$.

There have been many attempts to get rid of this unwanted part, that exists also in 
crystals; for example a very interesting observation is that most of the negative 
eigenvalues at $T<T_{c}$ are not related to existence of two different minima, i.e. if we 
start the minimization of $H$ moving in one of the unstable directions we arrives always 
the same minimum \cite{water}.  It has also been suggested that the distinction among 
localized and extended normal modes is relevant.  In any case we can assume that the INM 
provide a good approximation to the spectrum of the Hessian of the free energy, and if a 
small fraction of eigenvalues is removed, we find the expected behaviour with the fraction 
$f$ of negative eigenvalues dropping nearly to zero at $T_{c}$.
\begin{center} \begin{figure}[bt] {\epsfysize=8cm       \epsffile{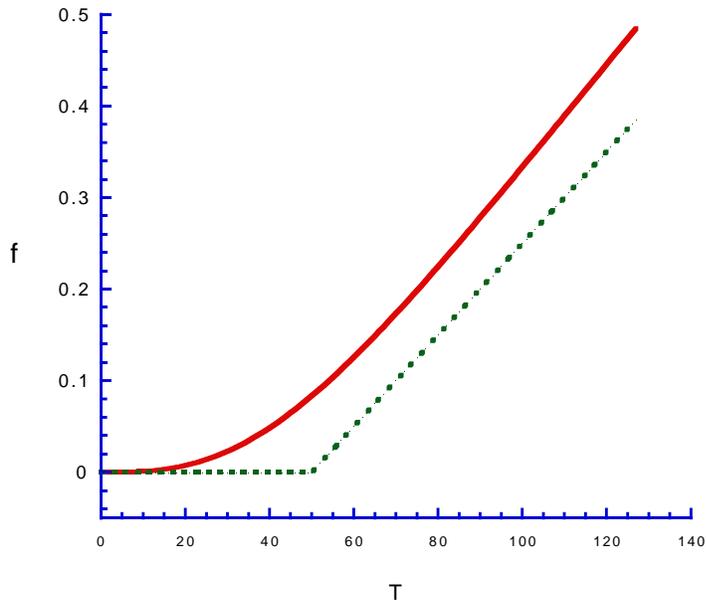}
    \caption{The qualitative behaviour of the fraction of negative eigenvalues ($f$) in a glass
    for the saddle normal modes (dotted line), for the 
    instantaneous normal modes (full line). In the case of saddles $f$ seems to vanish 
    very near to the dynamical transition $T_{c}$.
    }\label{F}
    }\end{figure}\end{center}

\subsection{The importance of being a saddle}

The study of saddle points and of the spectrum of the small fluctuation nearby, i.d. SNM 
(saddle normal modes) is a recent achievement. The relevance of saddles has been stressed 
in \cite{C} and the properties of the spectrum have been analyzed in \cite{ab}.

A saddle is a generic stationary point of the Hamiltonian and it is a solution
of the equations
\be
{\partial H \over \partial x_{i} } \equiv \sum_{k}V'(x_{i}-x_{k})=0 \ .
\ee
In other words the force on all the particles is equal to zero on a saddle. The following 
specializations are available: when all the eigenvalues are positive the saddle is a 
minimum and when all the eigenvalues are negative it becomes a maximum.

In glasses the number of saddles is exponentially large when the volume goes to infinity.  
For each configuration $x_{i}$ we can associate an inherent saddle, i.e. a saddle that is 
the nearest one to that particular configuration (the details of this definition depend
on the precise definition of distance, however it has the nice feature that the quest for 
saddles near to a given configuration can be implemented numerically in a meaningful 
way).

At a given temperature we have a probability distribution of the configuration  and this 
naturally induces a probability distribution on the saddles. As far as the inherent saddles should 
be not {\sl too} distant from the the original configuration the SNM spectrum should be not {\sl too 
different} from the INM spectrum. However this is one of the cases where a small difference 
matters. 

\begin{center} \begin{figure}[bt] {\epsfysize=8cm       \epsffile{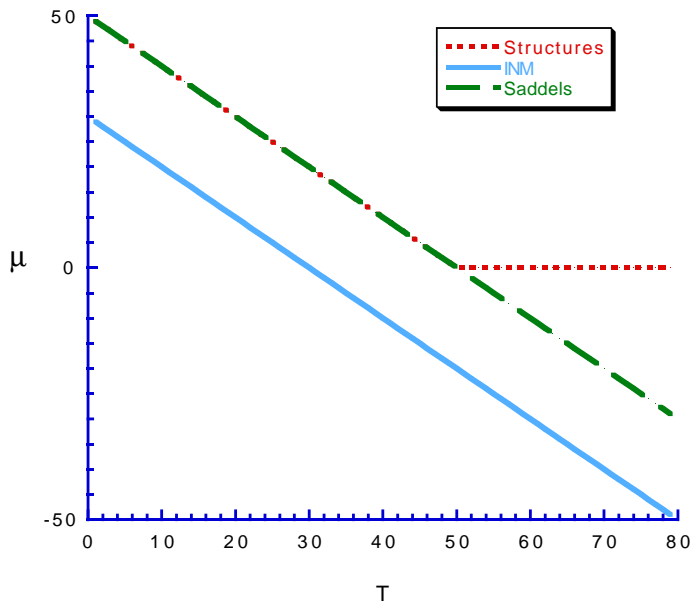}
    \caption{The qualitative behaviour of $\mu$ as function of the temperature
    in mean field approximation for the saddle normal modes (dotted line), for the 
    instantaneous normal modes (full line) and for the inherent structures normal modes (dashed line).
    }\label{MU}
    }\end{figure}\end{center}

In mean field theory the analytic computation of the SNM spectrum can be done: we find 
that if we start from a configuration below $T_{c}$, the nearby configuration is a 
minimum and the spectrum has a gap; this gap moves toward zero when we approach $T_{c}$ 
and in the region where $T>T_{c}$ the fraction of negative modes becomes different from 
zero, i.e we recover the behaviour described in figure (\ref{MU}).

In a surprising way we find numerically the same effect for a fluid: the qualitative 
behaviour of the fraction $f$ of negative SNM as function of the temperature is shown in 
fig.  (\ref{F}) (for comparison we shown also the fraction of negative INM).  A similar 
conclusion arise also if we plot the average fraction of negative SNM ($f$) as function of 
the saddle energy.  The quantity $f$ vanishes just at the energy that correspond to the 
dynamical transition $T_{c}$.

The smart reader would now observe that if our starting configuration is taken in the act 
of jumping from one minimum to an other minimum, it should be near to a saddle with a 
negative eigenvalue and therefore the fraction of SNM is never zero, also at low 
temperature.  However it is quite plausible that these activated processes are quite rare 
and the fraction of negative negative eigenvalues is not zero below $T_{c}$ but it maybe a 
rather small number (e.g. $10^{-6}$), which certainly do not affect the overall picture.

The conclusion is that the spectrum of the SNM has just the properties that we supposed 
for the eigenvalues of the Hessian of the free energy: negative eigenvalues appear in a significative way 
only above the critical temperature $T_{c}$. At temperature smaller than $T_{c}$ the 
saddles become minima and the SNM coincide with the inherent structure normal modes of the 
appropriate energy density.

At temperature higher than $T_{c}$ the properties of the saddles are definitely different 
form those of the the inherent structures (i.e. the minima of the Hamiltonian that are 
the nearest to an equilibrium configurations), however when we decrease the temperature 
toward $T_{c}$, the inherent saddles loose the negative part of the spectrum and they 
start to look like minima and their properties becomes the same of those of the 
inherent structures. Indeed at temperatures lesser than $T_{c }$ the saddles are minima and
the inherent saddles coincide with the inherent structures.

\section{The Boson peak}

The conclusions of the previous section provide us a natural interpretation for the Boson 
peak. In this section hand waving physical arguments will be presented, while in the next section we will 
present a precise and technical analysis that supports the conclusions of this section.

The reader must have noticed that the computations of the spectrum we alluded in the 
previous section were not done for fluids but only for simplified mode coupling models. 
This fact has the sad consequence that phonons are absent from the spectrum. 
In these models the spectral density has a very simple behaviour: it is a semicircle law 
starting from a value $\mu(T)$. The main results we have described in the previous section are the 
following (see figure \ref{MU}):
\begin{itemize}
     \item
    For the inherent structures $\mu(T)$ is definitively positive below $T_c$, it goes to 
    zero at $T_{c}$  and it remains 
    zero in the fluid region above $T_c$.
   \item
    For instantaneous normal modes $\mu_(T)$ is negative in the fluid phase and remains 
    negative  at $T_{c}$ (it may become positive at definitively lower temperatures) and it 
    reach the $\mu(T)$ of the inherent structures when we arrive at zero temperature.
    \item
    For the inherent saddles $\mu(T)$ is negative in the fluid case, it goes to zero at 
    $T_{c}$ and coincide with that of the inherent structures below $T_{c}$.
\end{itemize}
    
A further piece of information is important.  We have already remarked that if we cool the 
system very fast below $T_{c}$ in the ideal mean field model the system is frozen in 
a metastable state.  It can be shown (as a consequence of the principle of marginal 
stability) that in this case the function $\mu(T)$ of the inherent structures (in this 
case it coincides with that of the inherent saddles) is equal to zero; on the other hand, 
if we suppose that the system is able to cross barrier of order $\Delta N$, the quantity 
$\mu$ becomes a continuos positive function of $\Delta$, vanishing at $\Delta=0$.  It is 
reasonable to assume that the behaviour of real systems is quite similar to that of the 
mean field theory with a finite $\Delta$, that increase logarithmically when we decrease 
the cooling rate.

In other words  at low temperature at equilibrium the density of states does not touch 
zero, however it arrives up to zero in the ultrafast quenched state; in the case of slow 
cooling the spectrum extend up to zero with a small gap. This physical picture implies 
that  a glassy system has  an anomalously high density of vibrational excitations at low energies in the 
low temperature region. It is natural to assume that this excess of low energy vibrations 
\footnote{These states are also connected with dynamic heterogeneities in cooling, but 
this relation cannot be discussed here for reasons of space.} is connected to the Boson 
peak. 

In order to substantiate this claim we should consider systems that have phonons and prove 
that  there is a Boson peak with all the correct physical properties.  In the best 
case one should be able to compute the properties of the Boson peak in a quantitative way.  
A microscopical approach which tries to obtain these results is described in the next 
section.

\section{A microscopic approach}

We come back to the original problem stated in the introduction. We would like to compute  the 
spectrum of the matrix $M$ associated to the quadratic form
\be
\frac12 \sum_{i,k}h(x_{i}-x_{k}) ( \phi_{i}-\phi_{k})^{2}\label{QUA} \ .
\ee
This matrix has interesting properties;  translational 
invariance predicts the presence of a phononic spectrum with density of states 
$\rho(z)$ that at small eigenvalue $z$ is greater or equal to $z^{2}$ in three 
dimensions.

The most interesting situation happens in the case of a non positive $h$.  Here negative 
eigenvalues are possible.  A very simple probabilistic argument shows that in the infinite 
volume limit a generic configuration of the $x$'s always has negative (may be localized) 
eigenvalues.

We can argue that there are two phases:
\begin{itemize}
    \item The {\sl  phonon} phase, where negative eigenvalues are localized. Here the spectral 
    density goes approximatively like $z^{1/2}$ at small (but not small) $z$.
    \item The {\sl saddle} phase where   negative eigenvalues are present and the spectral 
    density goes like a constant at small $z$.
\end{itemize}

However in a first approximation the density of localized eigenvalues is small.  If we 
neglect them, we can state that in the phonon phase negative eigenvalues are absent and 
the spectral density goes like $z^{2}$ at small $z$.
   
Let us consider the case where the function $h(x)$ depends on a parameter $g$ that we can 
tune in such a way that for $g>g_{c}$ we are in the phonon phase, while for $g<g_{c}$ we 
are in the saddle phase.  We would like to show that a Boson peak is present in the phonon 
case in the region of $g$ near to $g_{c}$.  In other words, if we consider a matrix of the 
form given in eq. (\ref{QUA}), we expect that when this matrix is loosing its stability (i.e. we are near 
to $g_{c}$) a Boson peak appear.  The Boson peak is therefore an universal feature of 
random matrices near the phonon-saddle transition and its properties should not depends 
too much on the microscopic details.
    
In order to prove this statement we have to make a microscopic computation of the spectrum 
and of the eigenvectors and to show that near $g_{c}$ a Boson peak is present.
The simplest way to reach this goal is to extend the CPA (coherent potential approximation) approach, 
(that works very well on the lattice) to off-lattice models. 

Usually in the CPA approach the density of states is  well approximated, however localized 
states are absent and one  finds a zero density of states at places where the density of 
states is very small. This drawback of the CPA is useful in our case because it gives an 
exactly zero density of states in the phonon phase at negative eigenvalues ($z$).

The generalization of the CPA to our case is not so simple and it has been obtained after 
a long series of works \cite{LONG,BOSE}. Let me be rather sketchy on this point. The crucial step consists in 
introducing (as usual) the resolvent
\be
R(j,k|z)\equiv \(({1 \over z -M }\))_{j,k} \ .
\ee
The Green function is the average of the resolvent, i.e.
\be
G(p,z)=\overline{\sum_{j,k} R(j,k|z) \exp(i (x_{j}-x_{k})} \ .
\ee
The Green function is related to the usual structure function by the relation:
\be
S(p,z)\propto p^{2}  z^{-1} \mbox{Im} \ G(p^{2},z^{2}+i 0^{+})
\ee

With some effort \cite{BOSE} one can derive an approximate integral equation for the resolvent which
looks like
\bea
G(p,z)^{-1}=G_{0}(p,z)^{-1}- \Sigma(p,z)\ , \\
\Sigma(p,z)=\int d^{3}q W(p,q)G(q,z) \ , 
\eea
with an appropriate definition of the {\sl bare} Green function $G_{0}(p,z)$ and of the 
vertex $W(p,q)$. These equations are similar to those  derived using a completely 
different  approach in the framework of the mode coupling theory \cite{MC}, however the form of the 
vertex is rather different and this leads to subtle differences that cannot be discussed 
here.

A detailed computation \cite{BOSE} shows that the previous  integral equations do have a transition 
from a phonon phase to a saddle phase and that a Boson peak is present  near the transition in 
the phonon phase. The discussion of the the comparison with scattering experiments and their
relevance in the framework of the glass transition cannot be done here for reasons of space and 
can be found in \cite{BOSE}.

We can this conclude that a Boson peak is a generic property of translational invariant 
matrices which are near the phonon-saddle transition and this phenomenon is not restricted 
to glasses.  On the contrary it is the duty of the theory of glasses to explain why 
glasses at low temperature are near to the phonon-saddle transition and modern theories of 
glasses have a clear explanation for this phenomenon.

\section*{Acknowledgements}
I have the pleasure to thank all the people which have worked with me in the subject of 
random matrices, i.e.  Andrea Cavagna, Barbara Coluzzi, Irene Giardina, Tomas Grigera, 
V\'{\i}ctor Mart\'{\i}n-Mayor, Marc M\'ezard, Paolo Verrocchio and Tony Zee.

\end{document}